# Human-centred Test and Evaluation of Military AI

Session at Responsible AI in the Military Domain, Seoul Korea, 9-10 Sep 2024

*Authors*: DR David Helmer[1], MR Michael Boardman[2], DR S. Kate Conroy3, LTCOL Adam J. Hepworth[4] and MR Manoj Harjani[5]

[1] John Hopkins University, Applied Physics Laboratory, USA

[2] Defence Science and Technology Laboratory, UK

[3] Queensland University of Technology, Australia

[4] Australian Army

[5] RSIS Nanyang Technological University, Singapore

## Summary

This session ' highlights the need for and provides understanding of the challenges associated with human-machine teaming (HMT) test and evaluation (T&E) in defense, particularly measurement and testing protocols. The session will equip attendees to lead productive engagements in their respective communities, with a particular focus on bridging the gap between the technical, operational and policy communities

The REAIM Call to Action (2023) advocates for thorough research, testing, and assurance before adopting AI in the military domain to avoid inadvertent harms. The purpose of this session is to provide the REAIM community context on test and evaluation (T&E) for human-machine teams (HMT). Traditionally there has been extensive discussion on testing of AI algorithms, but less regarding HMT. There are a number of practical and technical challenges associated with such T&E – this session will lay out these issues, associated questions and implications in a manner accessible to the broader community (technical and nontechnical). Greater understanding of T&E will help to drive responsible AI policy development, as well as acquisition and deployment of these systems. This session will better enable treatment of these issues and will motivate investment in resolving these challenges across the community.

*Guiding questions:*

1. How can we incorporate human considerations into a reliable T&E process and develop a skilled workforce to support this integration?

2. What tests and metrics are needed to evaluate HMTs and how do they differ from those for AI algorithms?

3. How can we best implement responsible AI for HMTs given the limitations of current T&E capabilities?

# Full Summary

| **Background & Objectives** |
|---|
| **Background** <br> The Political Declaration on the responsible use military of AI and autonomy (Nov, 2023) calls for the endorsing states to develop appropriate measures to ensure responsible AI at relevant stages of the AI lifecycle. The REAIM Call to Action (Feb 2023) advocates for thorough research, testing, and assurance before adopting AI in the military domain to avoid inadvertent harms. The purpose of this session was to provide the REAIM community context on test and evaluation (T&E) for AI systems that will interact with humans (encompassing nearly all systems). <br><br> Traditionally there has been extensive discussion on test and evaluation (T&E) of AI algorithms, but much less discussion regarding a more wholistic approach to T&E that incorporates consideration of the human component (such as in human-AI teams). There are a number of practical and technical challenges associated with such T&E – this session sought to lay out these issues, associated questions and implications in a manner accessible to the broader community (technical and nontechnical). As humans are considered central to the responsible use of AI and remain accountable for the actions of such systems, greater consideration of the human component within T&E will help to drive responsible AI (RAI) policy development, acquisition, and deployment of these systems. This session will better enable treatment of these issues and will motivate investment in resolving these challenges across the community. |

T&E of AI-enabled systems is often poorly understood by policymakers. As the international community shifts its focus from making RAI commitments to attempting to operationalize them, it is essential that effective strategies are in place to ensure the consistency of actions in accordance with RAI principles. T&E is how one ensures AI-enabled systems follow policy and as such its role in RAI is fundamental.

In real-world applications, few systems operate in a vacuum. Most often, AI-enabled systems will operate with human users or teammates in a Human-Machine Team (HMT). The effective incorporation of humans is critical to responsible application of AI in the military domain, though the manner in which the human is incorporated may vary widely. The teaming constructs employed in HMTs add another layer of complexity to the T&E landscape.

In HMTs, the human will generally be expected to fill a role that ensures acceptable system (human with machine) behavior for both practical implementation and operational reasons, as well as and ethical and policy alignment. In the commercial world, this expectation is not always met, particularly when systems are designed with sole focus on optimizing the algorithmic performance. To enforce reliability, accountability, and transparency in practice, it is imperative that we perform T&E with representation of the human.

Despite the research on T&E of safety-critical systems, some of which employ AI, and centuries of research on human behavior, significant gaps remain in our ability to perform reliable T&E for HMT. Operators consistently surprise designers with use (or disuse) of fielded capabilities even without embedded AI. With AI, T&E approaches typically rely on massive numbers of tests due to the complex scenarios in which they are usually deployed. However, this sort of scaling is fundamentally incompatible with human users – one might at best get tens of test participants for most systems.

Further expanding T&E difficulty, it is well known that user experience and interactions can dramatically impact human behavior, understanding of the system and ability to interact with it (and therefore system performance). For example, a system with 95% AI reliability with which a human struggles to remediate the gaps may in actuality be significantly less reliable than one with 90% algorithm reliability but a more effective HMT. Capturing this performance trade-off remains an open problem. If we cannot reliably capture expected system performance in T&E, we will struggle to fully implement RAI principles in the military domain. In this session, attendees will be provided with a comprehensive understanding of the current challenges, state of the art, common practices, and potential paths forward in a manner designed to bridge the gap between the technical and nontechnical communities, which can foster informed dialogue on this issue.

**Guiding Questions:**
1. How can we incorporate human considerations into a reliable T&E process and develop a skilled workforce to support this integration?
2. What tests and metrics are needed to evaluate HMTs and how do they differ from those for AI algorithms?
3. How can we best implement responsible AI for HMTs given the limitations of current T&E capabilities?

**Objectives**
1. Highlight the need for and provide understanding of the challenges associated with HMT T&E in defence particularly measurement and testing protocols.
2. Equip attendees to lead productive engagements in their respective communities, with a particular focus on bridging the gap between the technical, operational and policymaker

**Main Discussion Points**

**David Helmer PhD (JHU/APL)**

1. The community faces fundamental gaps in predicting real-world performance of AI-enabled capabilities due not only to challenges in AI test and evaluation, but an inability to predict the effect of interactions of humans with those systems
    1.1. This applies to essentially all capabilities – when we talk about autonomous or semi-autonomous "systems", the "system" should include the algorithm and/or platform but also the humans whose actions affect performance
2. Fundamental gaps exist in understanding and predicting human behavior even with simple systems; this extends to more complex AI-enabled systems
3. Typical AI performance measures focused on algorithmic performance are inadequate, as they do not reflect sufficient understanding of realized performance in operational context, which includes outcomes that are impacted by interactions with humans (which can both improve and degrade performance).
    3.1. This performance impact again emphasizes that for Responsible AI deployment, we must take a system-of-systems view, where the human is considered as part of the system
    3.2. Many capabilities assume that a human will intervene to fix machine errors, but the community generally does not currently effectively test to validate this assumption. And the possibility of human negation of a correct machine choice is often unaccounted for
4. The community must understand gaps in evaluation of these systems and close them technically while understanding implications to risk evaluation and acceptance
    4.1. Some example problems include the fact that most AI requires many test points to evaluate, while humans can run only a small number of cases. Additionally, this testing is traditionally executed late in development, which makes change implementation difficult. Challenges exist for adjusting timelines, and there are no current demonstrated

> abilities to mitigate the limited test case problem nor to robustly understand the resultant unknown risk due to sample bias, limited data sets, etc.
>
> 4.2. This is not just a technical challenge. It is critical that the legal and policy communities understand the constraints and possibilities to drive appropriate decision-making

5. The goals and needs of this test, evaluation, verification, and validation (TEVV) process are not fundamentally changed for AI in comparison with legacy systems – in many ways these problems already exist for systems today. Challenges may be exacerbated, but the need to define and/or mitigate risk in TEVV is unchanged. The means to do so will in some cases need to evolve.

**Mr Michael Boardman (DSTL)**

6. Human Machine Teaming approaches and the onus of accountability resting on the human user within AI-based systems will demand an approach to TEVV that isn't just technologically focused. It must include the user as part of the system, together with consideration of their training, mental models/understanding of the system and capability to meaningfully interact with the technological component of the system when needed.

7. TEVV will be an ongoing process across the lifecycle of the system – where context of use, operational environment etc change TEVV must reoccur (potentially rapidly and in theatre).

8. Means of communicating the results of TEVV to those using and making decisions regarding the use of AI based systems will be key in informing risk based decisions regarding use.

**DR S. Kate Conroy (QUT)**

9. The [REAIM 2024 Blueprint for Action](#) ('Blueprint'):

    *10. Commit to engaging in further discussions and to promoting dialogue on developing measures to ensure responsible AI in the military domain at the national, regional and international level, including through*

> *international normative frameworks, rigorous testing and evaluation (T&E) protocols, comprehensive verification, validation and accreditation (VV&A) processes, robust national oversight mechanisms, continuous monitoring processes, comprehensive training programs, exercises, enhanced cyber security and clear accountability frameworks;*

10. Digital engineering is preparing test and evaluation environments for emerging AI-enabled systems.
11. Test and evaluation protocols and VV&A processes need to test AI-enabled systems with human operators to ensure they can be used responsibly in accordance with the Blueprint

**Lieutenant Colonel Adam J. Hepworth, PhD (Australian Army)**

12. Traditional military governance and assurance models, which are often prescriptive and input-based, may not be adequate for AI's dynamic and unpredictable nature. An outcome-focused approach is necessary to ensure AI systems achieve desired results while adhering to ethical and legal standards.
13. Confidence-building measures, such as sharing AI performance data, are crucial for collaboration and mutual understanding.
14. Often there exists a gap between ethical discourse and practical military operations. TEVV frameworks are essential but must be grounded in operational realities to effectively support human-centric decision-making.
15. AI-enabled systems should operate contextually and predictably within defined situations. Rigorous TEVV processes are essential to assuring human military commanders and operators of system employment.
16. Additionally, robust governance frameworks must be in place at every level, from design to deployment, ensuring that the autonomy granted to AI systems is carefully managed and aligned with operational objectives.

**MR Manoj Harjani (RSIS Nanyang Technological University)**

17. There are potential challenges arising from standards for human-centred T&E being politicized - i.e., it becomes less about the standards and more about who has been involved in developing them.

18. We also need to consider the organisational politics within militaries that will affect T&E implementation - the standards and practices that eventually are adopted will be shaped by the interest group that wins out.

19. Anticipatory methods are one possible tool for militaries to navigate the challenges associated with the politics surrounding T&E implementation

**Audience**

1. Make-up
    1.1. 20% core work in T&E
    1.2. 80% expertise in Operationalization, Data practices, Intl law & policy, Diplomacy, Arms control, Human rights, Law, Defence policy, Regulation
2. Audience felt that human-centred test and evaluation of military AI is (>50% participants assent, little dissent):
    2.1. AI systems are tested with human operators in their intended context of use and anticipated deployment conditions ahead of use
    2.2. AI systems are evaluated for their effectiveness with human operators
    2.3. AI systems deployed in their intended use comply with international humanitarian law.
3. Audience felt that human-centred test and evaluation of military AI POSSIBLY is (more participants assent than dissent)
    3.1. Human feedback drives product updates throughout the AI lifecycle
    3.2. AI systems are assessed for human impacts to inform risk management
    3.3. Human factors of AI systems affect acquisition decision making
    3.4. AI systems are designed to augment human decision making
    3.5. AI systems are designed to improve human agency and autonomy
4. Audience felt that human-centred test and evaluation of military AI MAY OR MAY NOT be (same participants assent and dissent)
    4.1. AI systems are designed for the wellbeing of operators
    4.2. Stakeholder values are elicited in order to determine requirements for AI systems
    4.3. AI systems are designed to reduce human biases

5. Audience felt that human-centred test and evaluation of military AI is NOT (more participants dissent than assent)

    5.1. Human performance is measured when using AI-enabled systems

    5.2. Those responsible for AI systems need to prioritise reducing unintended harms that may be caused by use of these systems.

    5.3. AI systems are designed to improve justice outcomes for humanity

6. Audience wants to get out of the workshop:

    6.1. AI and relevance to asymmetry in conventional wars

    6.2. How international human rights law could inform

    6.3. How testing and eval contributes to legal reviews

    6.4. Why humans should be centered instead of ai

    6.5. Awareness raising on international law

    6.6. A broad perspective on how Nations are approaching

    6.7. Awareness of human-centred T&E's importance

    6.8. Some concrete approaches of test and evaluation

7. Audience did not want to spend time on:

    7.1. Mathematics

    7.2. Too much conceptual discussions

    7.3. LAWS

    7.4. AI ethics principles

    7.5. Definitions

**Key Takeaways**

**Lieutenant Colonel Adam J. Hepworth, PhD (Australian Army)**

1. The REAIM Blueprint noted that AI applications should be ethical and human-centric and that humans must remain responsible and accountable for their use and effects. Developing rigorous TEVV frameworks will contribute to robust oversight mechanisms.

2. Adopting a human-centric model of 'decision custodianship' where algorithms act as tools to augment human capacity needs further attention. Ensuring that the behaviour and outputs of AI-enabled systems are reliable and consistent will provide confidence for human military commanders and operators.

3. Developing cross-function teams is required to ensure an exhaustive approach to TEVV in military operations.

**MR Manoj Harjani (RSIS Nanyang Technological University)**

4. The need for dialogue between technologists and policymakers on human-centred T&E will be evergreen - but dialogue needs to be initiated with an objective in mind for it to be productive.
5. There will be a need to address potentially fragmented and siloed implementations of T&E across different countries' militaries - absolute consensus and complete interoperability is unrealistic, but there should be a focus on implementation according to standards that are more widely accepted.
6. How can T&E become a focal point for confidence-building among militaries?

**Remaining Challenges and Issues**

**DR S. Kate Conroy (QUT)**

1. Test and evaluation in the development of AI systems needs to involve human users and the development of AI systems may go on throughout the deployment of the AI system in agile, iterative AI development, iteration, updates and adaptation within contexts of operations
2. Traditional human-centred test and evaluation methods from human factors including objective measures (e.g. eye tracking) and subjective measures (e.g. self-reports) of fatigue, cognitive overload, attention, wellbeing etc.. need to be adapted for deployed AI systems that require ongoing monitoring and evaluation.
3. New standards need to be developed to help users of general purpose AI systems to capture qualities of these systems such as veracity, useability, productivity advantage, meaningful work etc..

**DR David Helmer (JHU APL)**

4. The community must acknowledge that essentially every AI-enabled system is impacted by a human, whether in a positive way (e.g. mitigating errors) or

   a negative. Test and evaluation in support of Responsible AI deployment MUST include the effect of the human to reflect operationally realized system performance. The language around AI-enabled systems should be shifted to inclusion of the human(s) as a component of the system
5. Standards and requirements supporting this adjusted definition are needed, as are metrics and means to evaluate them
   5.1. Effective modeling and simulation is highly desirable, but is not currently achievable in any validated manner.
   5.2. Development of test and evaluation that is achievable earlier in the process and also throughout system lifecycle is critical to support this evolution for a variety of reasons, including the issue of human scalability and impact on scale of achievable testing
6. Communication between technical and non-technical communities must be improved to ensure operators and policy-makers understand risk assumed by system use and to better inform research and development by technologists in the test and evaluation space

**Future Plans**

Collaboration to deliver to REAIM 2024 Blueprint for Action.